\documentclass[12pt]{article}
\usepackage{graphics}
\def\al{\alpha}
\def\be{\beta}
\def\ga{\gamma}
\def\de{\delta}
\def\ep{\epsilon}

\def\la{\lambda}
\def\th{\theta}

\def\Ga{\Gamma}
\def\De{\Delta}

\def\half{{1\over{2}}}
\def\pd{\partial}

\newcommand{\beq}{\begin{equation}}
\newcommand{\eeq}{\end{equation}}
\newcommand{\bea}{\begin{eqnarray*}}
\newcommand{\eea}{\end{eqnarray*}}
\newcommand{\beaq}{\begin{eqnarray}}
\newcommand{\eeaq}{\end{eqnarray}}
\begin{document}
\begin{flushright}\end{flushright}
\centerline{\Large \bf RG Flows from Super-Liouville Theory}
\vskip .5cm
\centerline{\Large \bf to Critical Ising Model}
\vskip 1cm
\centerline{\large Changrim Ahn\footnote{ahn@ewha.ac.kr},
Chanju Kim\footnote{cjkim@phya.snu.ac.kr},
Chaiho Rim\footnote{rim@mail.chonbuk.ac.kr},
and Al. B. Zamolodchikov\footnote{zamolod@lpm.univ-montp2.fr}}
\vskip 1cm
\centerline{\it$^{1}$Department of Physics, Ewha Womans University}
\centerline{\it Seoul 120-750, Korea}
\vskip .5cm
\centerline{\it $^{2}$ School of Physics, Seoul National University}
\centerline{\it Seoul 151-747, Korea}
\vskip .5cm
\centerline{\it $^{3}$ Department of Physics, Chonbuk National University}
\centerline{\it Chonju 561-756, Korea}
\vskip .5cm
\centerline{\it $^{4}$ Yukawa Institute of Theoretical Physics}
\centerline{\it Kyoto University, Kyoto 606-8502, Japan}
\centerline{\it and}
\centerline{\it Laboratoire de Physique Math\'ematique, 
Universit\'e Montpellier II}
\centerline{\it Pl.E.Bataillon, 34095 Montpellier, France}

\vskip 1cm
\centerline{\small PACS: 11.25.Hf, 11.55.Ds}
\vskip 2cm
\centerline{\bf Abstract}

We study an integrable deformation of the super-Liouville theory
which generates a RG flows to the critical Ising model as the IR
fixed point. 
This model turns out to be a supersymmetric sinh-Gordon model
with spontaneously broken $N=1$ supersymmetry.
The resulting massless Goldstino is the only stable on-shell
particle which controls the IR behaviours.
We propose the exact $S$-matrix of the Goldstino 
and compare associated thermodynamic Bethe ansatz equations with the 
quantization conditions derived from the reflection amplitudes
of the the super-Liouville theory to provide nonperturbative checks
for both the (NS) and the (R) sectors.

\newpage

\section{Introduction}
Integrable quantum field theories defined in two dimensions can be 
formally written as a UV conformal field theory (CFT) perturbed by
some relevant operator \cite{Zamol}.
Most well-known examples are the unitary minimal CFTs ${\cal M}_{p}$
perturbed by the least relevant field $\Phi_{1,3}$ whose action can
be written formally as follows:
\beq
{\cal M}A_{p}^{\pm}={\cal M}_{p}+\la\Phi_{1,3}.
\eeq
Here, the sign $\pm$ stands for the signature of the coefficient $\la$.
If the coefficient of the perturbation is negative,
the perturbed CFT ${\cal M}A_{p}^{-}$ is described by the factorized
scattering theory of massive particles called kinks.
More interesting is the case of ${\cal M}A_{p}^{+}$ which is shown to 
generate RG flows from the UV CFT ${\cal M}_{p}$ to ${\cal M}_{p-1}$.
This was first noticed in \cite{zamolii} by perturbative computation for
the case of $p>>1$ and was proved later by the thermodynamic Bethe 
ansatz (TBA) based on the $S$-matrix of the massless kinks \cite{alyosha}.

Among these, the RG flows from the tricritical Ising model (TIM) 
to the critical Ising model draws a particular interest since
the TIM is a super CFT while the Ising model is not \cite{alyoshaii}.
An analysis based on the Landau-Ginzburg potential in \cite{KMS} 
shows clearly how the RG flow can be understood.
The unperturbed TIM has a $\Phi^3$ superpotential, which is in components
${\bar\psi}\psi\phi+{1\over{8}}\phi^4$. 
The relevant perturbation, $\Phi_{1,3}$, is the top component of the
superfield $\Phi$ and preserves the supersymmetry.
This modifies the superpotential to 
${\bar\psi}\psi\phi+{1\over{2}}({1\over{2}}\phi^2+\la)^2$.
For $\la<0$ the ground state energy is zero, so supersymmetry is unbroken
and both boson and fermion become massive.
The $S$-matrix is non-diagonal and commutes with the supercharges 
\cite{zamoliii}.

With the positive coefficient $\la>0$, the superpotential generates
nonvanishing ground state energy and the supersymmetry becomes
spontaneously broken.
The bosonic field becomes massive, but the fermion stays massless
and plays the role of Goldstino.
In the IR limit one can integrate out the massive bosonic field to 
obtain the effective theory described by the Volkov-Akulov field
theory\cite{VolAku}
\beq
{\cal L}_{\rm VA}=-{1\over{2\pi}}(\psi{\bar\pd}\psi+{\bar\psi}\pd{\bar\psi})
-g(\psi{\pd}\psi)({\bar\psi}{\bar\pd}{\bar\psi})
+\cdots
\label{VAaction}
\eeq
where $\cdots$ include higher dimensional operators.

In this paper, we propose another RG flow where the supersymmetry is
spontaneously broken and the low energy effective action is described by
a Goldstino.
The model is another supersymmetric sinh-Gordon (SShG) model which
can be considered as a perturbed super-Liouville field theory (SLFT)
\cite{Alyosha}.
The ordinary SShG model is one of the simplest examples of 
a  $1+1$ dimensional 
integrable quantum field theory with $N=1$ supersymmetry \cite{SSG}. 
A generic lagrangian including one scalar superfield can be expressed
in terms of the component fields as
\beq
{\cal L}(\Phi)={1\over{8\pi}}(\pd_{a}\phi)^2
-{1\over{2\pi}}(\psi{\bar\pd}\psi+{\bar\psi}\pd{\bar\psi})
-{i\over{4\pi}}\psi{\bar\psi} W''(\phi)
+{1\over{32\pi}}\left[W'(\phi)\right]^2 .
\label{eqLag}
\eeq
The ordinary SShG model is a particular case of Eq.(\ref{eqLag}) 
with the superpotential
\beq
W(\phi)=-8\pi\mu\cosh(b\phi).
\label{eqSP}
\eeq
The SShG model and its imaginary coupling version ($b\to i\beta$), 
the supersymmetric sine-Gordon (SSG) model, are integrable since 
they can be mapped into an affine Toda theory based on the twisted
super-Lie algebra $C^{(2)}(2)$ \cite{Toda}.
This model preserves the supersymmetry and the boson and fermion
remain massive.
Exact factorized nondiagonal $S$-matrices have been obtained from
the integrability and on-shell supersymmetry in \cite{ShaWit,Ahn}.
This model is analogous to the TIM with $\la<0$.

Another SShG model, which is our main concern in this paper, is
defined by a slightly different superpotential, namely,
\beq
W(\phi)=-8\pi\mu\sinh(b\phi).
\label{SP}
\eeq
The supersymmetry and integrability are all preserved.
If we consider an imaginary coupling $b=i\be$, the two supersymmetric 
sine-Gordon models become equivalent since one can 
shift the scalar field by $\phi\to\phi+{\rm const}$.
However, with a real coupling $b$, the new SShG model shows 
the RG flows from the UV super-LFT to the critical Ising model at the IR.
With the superpotential Eq.(\ref{SP}), the ground state energy does not
vanish so that the supersymmetry is spontaneously broken.
While the bosonic field $\phi$ remains massive, the 
fermion field becomes massless and is identified with Goldstino.
Supersymmetry prohibits the quantum corrections from generating mass.
Meanwhile, the bosonic field $\phi$ is unstable and decays into the 
massless fermions.
After the massive bosonic field is integrated out, the low energy 
effective action is described by the Volkov-Akulov action 
Eq.(\ref{VAaction}).

Stable on-shell particle states of this model 
are composed of massless left- and right-moving fermions,
$\psi_{L}$ and $\psi_{R}$, respectively.
This model can be thought of as a perturbed super-LFT analogous to the 
perturbed TIM with $\la>0$.
The $S$-matrix between the $\psi_{L}$ and $\psi_{R}$ can be conjectured
from the unitarity and crossing symmetry as well as a perturbative
computation.
In this paper, we propose the $S$-matrix with the assumption of
strong-weak coupling duality.

Non-perturbative confirmation of the conjecture is provided by
the TBA analysis.
For the cases of perturbed rational CFTs, the UV limit of the TBA provides
the central charges and conformal dimensions for the UV CFTs.
For the perturbed SLFT, one can extract out 
an additional information, namely, the reflection amplitudes from
the TBA.
We analyze the UV behaviour of the TBA equations of the new
SShG model with the conjectured $S$-matrix and compare it with
the reflection amplitudes of the super-LFT.
Numerical agreement with very high accuracy will estabilish the 
correctness of the $S$-matrix.
We also provide the IR analysis of the TBA equations and relate them
to the IR action Eq.(\ref{VAaction}).


\section{$S$-matrix and TBA}

Without any mass degeneracy, the $S$-matrix of the new
SShG model is diagonal.
The only interaction term, $\psi_{L}\psi_{R}\sinh(b\phi)$, if expressed
with chiral fermions $\psi_{L}$ and $\psi_{R}$, gives 
trivial scattering between the two $\psi_{L}$'s (and
two $\psi_{R}$'s), i.e. 
\beq
S_{LL}(\th)=S_{RR}(\th)=-1.
\eeq

Nontrivial $S$-matrix arises between a $\psi_{L}$ and a $\psi_{R}$.
Since the particle is massless, the scattering amplitude satisfies
the crossing-unitarity relation,
\beq
S_{LR}(\theta)S_{LR}(\theta+i\pi)=1.
\eeq
This equation is solved by the CDD factor 
\beq
S_{LR}(\theta)={\sinh\theta-i\sin\pi B\over{\sinh\theta+i\sin\pi B}},
\label{SRL}
\eeq
where we fix the rapidity by choosing a scale $M$ in such a way that
the energy-momentum is given by (for the $\psi_{R}$) 
$E=P={M\over{2}}e^{\theta}$.
Apparently Eq.(\ref{SRL}) is not the unique CDD choice. 
It is the minimal CDD factor which contains the proper resonance pole 
in the s- and u-channels with the resonance mass $m^2=M^2 e^{-i\pi B}$.

Without any bootstrap procedure, we can not fix the location of the 
resonance pole. 
Our conjecture for the parameter $B$ is 
\beq
B(b)={b^2\over{1+b^2}}.
\label{param}
\eeq
This is consistent with perturbation theory upto the second order
and preserves the duality $b\to 1/b$ enjoyed by the ordinary SShG model.
The duality has root in its UV CFT, namely the super-LFT which is dual.
Since the new SShG model can be also considered as a perturbed super-LFT,
it is plausible to assume the duality in our case.
Subsequently, we will provide nonperturbative confirmation of the $S$-matrix.

For this purpose, we compute the effective central charge of the SShG model
using the TBA analysis.
It is straightforward to write down the TBA equations from the $S$-matrix.
\beaq
\ep_{L}(\th)&=&\half MRe^{\th}-\int_{-\infty}^{\infty}\varphi(\th-\th')
\ln\left(1+\eta e^{-\ep_{R}(\th')}\right){d\th'\over{2\pi}},\\
\ep_{R}(\th)&=&\half MRe^{-\th}-\int_{-\infty}^{\infty}\varphi(\th-\th')
\ln\left(1+\eta e^{-\ep_{L}(\th')}\right){d\th'\over{2\pi}},
\eeaq
where the parameter $\eta$ is either $+1$ for the the Neveu-Schwarz 
(NS) sector or $-1$ for the Ramond (R) sector
and the kernel, the logarithmic derivative of the $S$-matrix, is
given by
\beq
\varphi(\th)= {4\sin\pi B\cosh\th\over{\cosh 2\th-\cos 2\pi B}}.
\eeq
The effective central charge is given by
\beq
c_{\rm eff}(R)={3MR\over{2\pi^2}}\int_{-\infty}^{\infty}
\left[e^{\th}\ln\left(1+\eta e^{-\ep_{L}(\th)}\right)
+e^{-\th}\ln\left(1+\eta e^{-\ep_{R}(\th)}\right)\right]d\th.
\label{cencharge}
\eeq

This TBA equation can be solved analytically in the UV region $MR<<1$.
Here, the $c_{\rm eff}(R)$ has logarithmic corrections
of $1/\log(MR)^n$ as leading contributions and subleading power corrections.
In particular, the $R^2$ term in $c_{\rm eff}(R)$ can be interpreted as
the vacuum energy contribution.
The analysis gives
\beq
{\cal E}_{0}={M^2\over{8\sin(\pi B)}}
\label{vac}
\eeq
which is the same as that of the sinh-Gordon model.
This result is somewhat expected since the vacuum expectation value
of the interacting potential can be determined by the (NS) reflection
amplitude of the $N=1$ super-LFT, which is the same as that of the LFT.
To compare the TBA result with the reflection amplitude, one needs
a relation between the dimensionful parameter $\mu$ and the mass scale
parameter $M$ for the SShG model.
We conjecture that this is the same as that of the ordinary SShG model 
given in \cite{ssgmass},  
\beq
{\pi\over{2}}\mu b^2\ga\left({1+b^2\over{2}}\right)=
\left[{M\over{8}}{{\pi B}\over{\sin \pi B}}\right]^{1+b^2}
\label{ssgmmu}
\eeq
with $\ga(x)=\Ga(x)/\Ga(1-x)$.
These conjectures will be confirmed by numerical analysis of the TBA 
equations in sect.4.

\section{Reflection Amplitudes and Quantization Condition}

The SShG model can be considered as a perturbed super-LFT 
whose lagrangian is given by
\beq
{\cal L}_{\rm SL}={1\over{8\pi}}(\pd_{a}\phi)^2
- \frac{1}{2\pi}(\bar\psi \pd \bar\psi + \psi \bar\pd \psi)
+i\mu b^2 \psi \bar\psi e^{b\phi} 
+{\pi\mu^2 b^2\over{2}} e^{2b\phi}. 
\label{actionslft}
\eeq
With the background charge $Q$
\beq
Q=b+1/b.
\eeq
This model is a CFT with the central charge
\beq
c_{SL}={3\over{2}}(1+2Q^2)
\eeq
and primary fields in the (NS) and (R) sectors.
A (NS) primary field $e^{\al\phi}$ has
dimension
\beq
\De_{\al}={1\over{2}}\al(Q-\al)
\eeq
and becomes degenerate with $e^{(Q-\al)\phi}$.
The two-point functions of the primary fields give 
the reflection amplitudes \cite{Rash,Pogho}. 
For the (NS) field, it is given by
\beq
S_{NS}(P)=-\left(\frac{\pi \mu }{2}\gamma \left(\frac{1+b^2}{2}\right)
\right)^{-\frac{2iP}{b}}
\frac{\Gamma (1+iPb)\Gamma\left(1+\frac{iP}{b}\right)}
{\Gamma (1-iPb)\Gamma\left(1-\frac{iP}{b}\right)}.
\label{refns}
\eeq
Similarly, for a (R) field $\sigma ^{(\epsilon)}e^{\al\phi}$ 
the reflection amplitude is given by
\beq
S_R(P)=\left(\frac{\pi \mu }{2}\gamma \left(\frac{1+b^2}{2}\right)
\right)^{-\frac{2iP}{b}}
\frac{\Gamma \left(\frac{1}{2}+iPb\right)
\Gamma \left(\frac{1}{2}+\frac{iP}{b}\right)}
{\Gamma \left(\frac{1}{2}-iPb\right)\Gamma
\left(\frac{1}{2}-\frac{iP}{b}\right)}.
\label{refr}
\eeq

To derive quantization conditions, one can consider
the super-LFT acting on the space of states 
\beq
{\cal A}_0={\cal L}_2(-\infty<\phi_0<\infty,\psi_0)\otimes{\cal F}
\label{sspace}
\eeq
where the fermionic zero-mode appears only for the (R) sector 
and ${\cal F}$ is the Fock space of bosonic and fermionic oscillators.
The appearance of bosonic and fermionic zero-modes in Eq.(\ref{sspace})
is well-known from the super-CFT results.
In the (NS) sector, there is no fermionic zero-mode since the fermion
field satisfies the anti-periodic boundary condition while it appears
in the (R) sector with perodic one.
The primary state $v_{P}$ 
can be expressed by a wave functional $\Psi_{v_{P}}[\phi(x_1)]$
which can be expanded in the asymptotic limit $\phi_0\to\infty$ as
\beq
\Psi_{v_{P}}[\phi(x_1)]\sim e^{iP\phi_0}+S(P)e^{-iP\phi_0}.
\eeq
The amplitude $S(P)$ is either $S_{NS}(P)$ or $S_{R}(P)$ depending on
the sector.

The ordinary SShG model defined by Eq.(\ref{eqSP}) can be considered as the
super-LFT (\ref{actionslft}) perturbed by 
\beq
\Phi_{\rm pert}=i\mu b^2\psi{\bar\psi}e^{-b\phi}+{\pi\mu^2 b^2\over{2}} 
e^{-2b\phi}.
\eeq
In the wave functional interpretation, the perturbing potential provides
another potential wall which confines the wave functional.
This leads to the quantization condition for the momentum and the
energy of the system in the limit that the size of the cylinder $R$  
goes to $0$.
The quantization condition and comparison with the TBA  based on the
nondiagonal $S$-matrix of the ordinary SShG model have been worked out
in \cite{AKR}.

The new SShG model, being considered as another perturbed super-LFT by
\beq
\Phi_{\rm pert}=-i\mu b^2\psi{\bar\psi}e^{-b\phi}+
{\pi\mu^2 b^2\over{2}} e^{-2b\phi},
\label{actionpert}
\eeq
can be analyzed in the same way.
One can obtain the quantization condition of $P$ for the (NS) sector, 
\beq \label{squantize}
\de_{\rm NS}(P)=\pi+2QP\ln{R\over{2\pi}},
\eeq
where $\de_{\rm NS}(P)$ is the phase factor of (NS) reflection amplitudes.
Similary, the quantization condition for the (R) sector becomes
\beq
\de_{\rm R}(P)={\pi\over{2}}+2QP\ln{R\over{2\pi}}.
\eeq
Notice that the main difference arises from the extra $-1$ factor in
front of the perturbing potential in Eq.(\ref{actionpert}).
Both conditions are invariant under $b\to 1/b$.

In terms of the quantized momentum $P$, 
the effective central charge is given by
\beq
c_{\rm eff}(R)=\left\{
\begin{array}{ll}
{3\over{2}}-12P^2+{6\over{\pi}}R^2{\cal E}_0&{\rm (NS)}\\
-12P^2+{6\over{\pi}}R^2{\cal E}_0&{\rm (R)}
\end{array}
\right.
\label{superceff}
\eeq
where we added the vacuum energy ${\cal E}_{0}$ to compare the 
same ground-state energy.

This quantization condition can be solved iteratively
by expanding $\de(P)$ in powers of $P$ and be compared with
the numerical TBA solutions:
\beq 
\label{ns_phase}
\de_{\rm K}(P) = \de_1^{\rm K} P + \de_3^{\rm K} P^3
+ \de_5^{\rm K} P^5 + \cdots
\eeq
where K stands for either NS or R.
Explicitly, the coefficients for the (NS) are given by
\beaq
\de_1^{\rm NS} &=& -2\left\{
\frac1b \ln\left[{\frac{\pi\mu}2} \ga\left(\frac{1+b^2}2\right)\right]
+\ga_{\rm E}Q   \right\} \nonumber\\
\de_3^{\rm NS}&=&\frac23\zeta(3)\left(b^3+\frac1{b^3}\right)\nonumber\\
\de_5^{\rm NS} &=& -\frac25\zeta(5) \left( b^5 + \frac1{b^5} \right)
\eeaq
and, for the (R),
\beaq
\de_1^{\rm R} &=& -2\left\{
\frac1b \ln\left[{\frac{\pi\mu}2} \ga\left(\frac{1+b^2}2\right)\right]
 + (\ga_{\rm E} + 2\ln2)Q   
\right\} \nonumber \\
\de_3^{\rm R}&=&-\frac13\psi^{(2)}(\frac12)\left(b^3+\frac1{b^3}\right)
            \nonumber\\
\de_5^{\rm R} &=& \frac1{60}\psi^{(4)}(\frac12) 
\left( b^5 + \frac1{b^5} \right).
\eeaq

\section{TBA analysis}

To derive the coefficients $\delta$'s from the TBA equations,
we derive the momentum $P$ as a function of $R$
from the scaling function $c_{\rm eff}(R)$ 
and compare with the quantization conditions to determine the coeffcients.
In Tables 1 and 2, we show the first three coefficients in
the expansion in powers of $P$ obtained by numerical analysis 
and compare with the corresponding coefficients from the reflection
amplitudes Eqs.(\ref{refns}) and (\ref{refr}) with $M=2$.
\begin{table}
\begin{tabular}{||c||c|c||c|c||c|c||} \hline
\rule[-.4cm]{0cm}{1.cm}B   & $\de^{\rm NS(TBA)}_1$ & $\de^{\rm NS(RA)}_1$ &
$\de^{\rm NS(TBA)}_3$ & $\de^{\rm NS(RA)}_3$ &
$\de^{\rm NS(TBA)}_5$ & $\de^{\rm NS(RA)}_5$ \\ \hline
0.3 & 0.276167   &  0.276167 &  3.08111  &   3.08111 & --3.49936
 & --3.49933 \\
0.35 & 0.823499  &  0.823499 &  2.34480 &   2.34480 & --2.03777
 & --2.03774 \\
0.4 & 1.17240    &  1.17240 &  1.90842 &   1.90842  &  --1.29352
 & --1.29349 \\
0.45 & 1.36725 &  1.36725 &  1.67590 &   1.67590 & --0.936165
 &  --0.936139 \\
0.5 & 1.42998 & 1.42998 &  1.60274 & 1.60274 & --0.829567
 & --0.829542 \\ \hline
\end{tabular}
\caption{First three coefficients of $\de^{\rm NS(TBA)}$ in the expansion
in powers of $P$ obtained by numerical analysis in comparison with
the corresponding $\de^{\rm NS(RA)}$.}
\end{table}

\begin{table}
\begin{tabular}{||c||c|c||c|c||c|c||} \hline
\rule[-.4cm]{0cm}{1.cm}B   & $\de^{\rm R(TBA)}_1$ & $\de^{\rm R(RA)}_1$ &
$\de^{\rm R(TBA)}_3$ & $\de^{\rm R(RA)}_3$ &
$\de^{\rm R(TBA)}_5$ & $\de^{\rm R(RA)}_5$ \\ \hline
0.3 & --5.77412   &  --5.77412 &  21.5677  &   21.5677 & --108.480
 & --108.479 \\
0.35 & --4.98943  &  --4.98943 &  16.4136 &   16.4136 & --63.1708
 & --63.1699 \\
0.4 & --4.48712    &  --4.48712 &  13.3590 &   13.3590  &  --40.0991
 & --40.0982 \\
0.45 & --4.20586 &  --4.20586 &  11.7313 &   11.7313 & --29.0212
 &  --29.0203 \\
0.5 & --4.11519 & --4.11519 &  11.2192 & 11.2192 & --25.7167
 & --25.7158 \\ \hline
\end{tabular}
\caption{First three coefficients of $\de^{\rm R(TBA)}$ in the expansion
in powers of $P$ obtained by numerical analysis in comparison with
the corresponding $\de^{\rm R(RA)}$.}
\end{table}

\begin{figure}
\rotatebox{0}{\resizebox{!}{15cm}{\scalebox{0.1}{%
{\includegraphics[0cm,0cm][22cm,22cm]{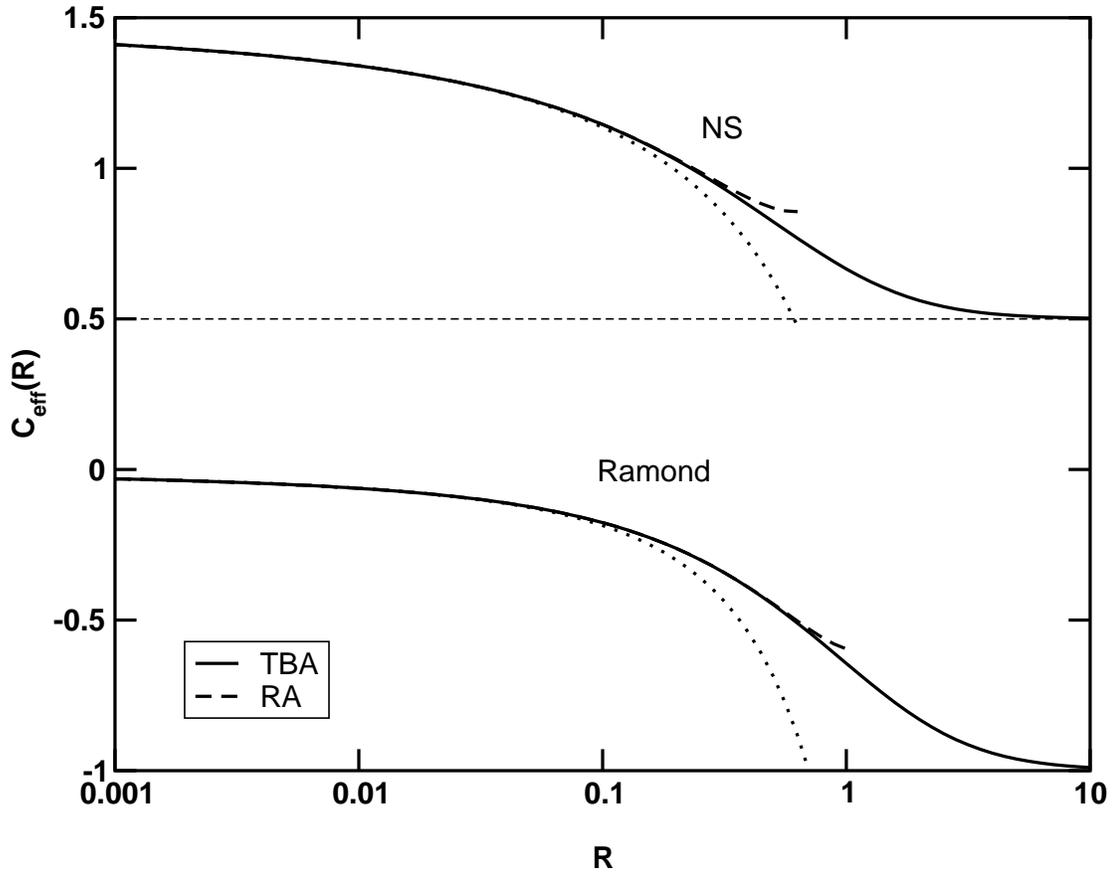}}}}}
\caption{Plot of $c_{\rm eff}$ for the (NS) and (R) sectors at $B=0.5$. }
\end{figure}

Our numerical analysis shows the consistency of the TBA equations
along with such conjectures as the scattering amplitude, the M-$\mu$
relation as well as the reflection amplitudes of the super-LFT
Eqs.(\ref{refns}) and (\ref{refr}). 
Also one can see from Fig.1 that the vacuum energy 
Eq.(\ref{vac}) improves the agreement with the TBA result
much better than without it (dotted lines) upto the range of $0.1<MR<1$.
This observation provides the validity of the conjectured vacuum energy.

We want to point out that a similar analysis for the ordinary 
SShG model in \cite{AKR}
could not give any non-perturbative confirmation for
the (R) reflection amplitude as well as the vacuum energy since
the scaling function for the (R) sector and the vacuum energy
vanish identically.
The new SShG model provides the unique ``experiment'' for these
quantities.

As suggested by numerical analysis, the SShG model flows into the Ising
model in the IR limit, $R\to\infty$.
The effective central charge in this limit is given by 
$c_{NS} = 1/2$ for the (NS) sector and 
$c_R = -1$ for the (R) sector where the Ramond vacuum with
the conformal dimension 1/16 is contributed.

In the IR limit $MR \ll 1$, 
the main contributions comes from the rapidity regions
where pseudo energy $\epsilon (\theta) \leq 1$. 
The asymptotic expansion can be obtained straightforwardly for
the (NS) and (R) sectors as follows:
\beaq
\label{c-NS}
c_{NS}&=&\frac12 +  \frac14 t + \frac14 t^2 
+ \left(\frac5{16} + {147\pi^2\over 400 }
{(2\cos 2\pi B+1)\over{\sin^2\pi B}} \right)t^3 + O(t^4)\\
\label{c-R}
c_{R}&=&-1+ t -2 t^2
+ \left(5 +  {4\pi \over 15} + {12\pi^2\over 25}
{(2\cos 2\pi B+1)\over{\sin^2\pi B}}\right)t^3 + O(t^4)
\eeaq
where 
\beq
t = {4\pi\sin\pi B\over{3(MR)^2}}.
\eeq

This IR behaviour can be described in terms of the Ising model 
with $T \bar T $ perturbation, Eq.(\ref{VAaction}). 
The perturbation contributes to $c_{eff}$ 
\beq
\label{c-TT}
c_{\rm eff} =c-12\left(\frac c {24}\right)^2 \alpha 
+ 12\left(\frac c {24}\right)^3 \alpha^2 + O(\alpha^3)
\eeq
with $\alpha = - 32 \pi^3 g /R^2 $ where $g$ is the coupling coefficient 
in Eq.(\ref{VAaction}).
Higher order term is ambiguous due to the UV regularization. 
Two results for the (NS) sector, Eqs.~(\ref{c-NS}) and (\ref{c-TT}),
are consistent upto order $\alpha ^2 $ and $t^2$ 
if we identify $c=1/2$ and $g = 2\sin\pi B/\pi^2 M^2 $. 
For the (R) sector, Eq.(\ref{c-R}) is consistent with
Eq.(\ref{c-TT}) when $c=-1$ and $g$ is the same as before.
Notice that the RG flow from the TIM to the Ising model is described by
Eq.(\ref{VAaction}) with $g=1/\pi^2 M^2$.

In summary, we have considered the SShG model with spontaneously broken
supersymmetry with a massless Goldstino which generates the RG flows
from the super-LFT to the Ising model for the (R) and (NS) sectors.
We propose a set of conjectures such as the $S$-matrix, $M$-$\mu$ 
relation, and the vacuum energy.
These conjectures are eventually justified by the independently drived
effective central charge based on reflection amplitudes of the super-LFT.

\section*{\bf Acknowledgement}

This work is supported in part by KOSEF R01-1999-00018 (C.A.,C.R.), 
KRF 2001-015-D00071 (C.A.) and
BK21 project of the Ministry of Education (C.K.).

\end{document}